# Emergence of Transient Domain Wall Skyrmions after Ultrafast Demagnetisation

Serban Lepadatu[1,*]

[1]*Jeremiah Horrocks Institute for Mathematics, Physics and Astronomy, University of Central Lancashire, Preston PR1 2HE, U.K.*

**Abstract**

It is known that ultrafast laser pulses can be used to deterministically switch magnetisation and create skyrmions, however the deterministic creation of a single Néel skyrmion after ultrafast demagnetisation remains an open question. Here we show domain wall skyrmions also emerge in systems with broken inversion symmetry after exposure to an ultrafast laser pulse, carrying an integer topological charge. Whilst domain wall skyrmions do not appear in the relaxed state due to quick thermal decay following an Arrhenius law, they play a key role in controlling the final skyrmion population through annihilations with skyrmions of opposite topological charge, with the resultant skyrmion states following a Poisson distribution. Using single-shot linearly polarised laser pulses, as well as a train of circularly polarised laser pulses, we show that when a high degree of disorder is created, the possibility of nucleating a single Néel skyrmion is accompanied by the possibility of nucleating a skyrmion with domain wall skyrmion pair, which results in a self-annihilation collapse.

[*] SLepadatu@uclan.ac.uk

The Dzyaloshinskii-Moriya interaction (DMI)[1,2] enables formation of a wide range of exciting magnetic objects which are topologically protected from collapsing into the background magnetisation state. The most widely studied topological object is the skyrmion[3], which has been observed both in materials with bulk DMI[4,5], and ultrathin films with interfacial DMI[6,7]. The topological charge is defined in Equation (1), and for a skyrmion it takes on unit values[8]. Such objects are intensely studied due to the ability to manipulate them with electrical currents through interfacial spin-orbit torques[9-11], bulk spin-transfer torques[12], as well as interfacial spin-transfer torques[13]. Skyrmions have also been proposed in antiferromagnetic materials[14] and could potentially be used as information carriers in antiferromagnetic spintronics[15]. Other related topological structures include anti-skyrmions[16,17], skyrmioniums[18], or more generally skyrmion bags[19]. Another type of object with unit topological charge is the domain wall (DW) skyrmion, which following initial theoretical studies[20-22], has been revisited recently[23]. The DW skyrmion is analogous to a vertical Bloch line which carries half-integer topological charge, but occurs in Néel domain walls as a 360º transverse rotation of magnetisation, and is stabilised by the interfacial DMI. Whilst such structures have not yet been observed experimentally, we show here they are ubiquitous in magnetisation recovery processes following ultrafast demagnetisation, appearing as transient topological objects which mediate interactions between skyrmions, and are thus essential for a full understanding of such processes.

$$Q_i = \frac{1}{4\pi}\int_A \mathbf{m}_i \cdot \left(\frac{\partial \mathbf{m}_i}{\partial x} \times \frac{\partial \mathbf{m}_i}{\partial y}\right)dxdy, \quad (i = A, B) \qquad (1)$$

Studies of ultrafast magnetisation dynamics have revealed the possibility of all-optical switching (AOS) of magnetisation, initially in ferrimagnetic GdFeCo[24], where a helicity dependence (HD-AOS) of magnetisation switching for a train of circularly polarised ultrafast laser pulses has been observed. Explanations of this helicity dependence have been given in terms of the effective field created due to the inverse Faraday effect[24,25], as well as magnetic circular dichroism[26]. On the other hand, it has also been shown heating alone due to the laser pulse can also result in deterministic switching of magnetisation without a helicity dependence[27,28], where the sample passes through a transient ferromagnetic-like state. HD-AOS has now been observed in a wide range ferromagnetic systems, alloys and multilayers[29,30], where deterministic switching of magnetisation was observed for a train of circularly polarised laser pulses. Additionally, ultrafast magnetisation switching has also been demonstrated using electronic heat currents[31,32], and deterministic switching using single-shot linearly polarised laser pulses was demonstrated in a synthetic ferrimagnetic racetrack[33]. The possibility of creating skyrmions in magnetic thin films was also investigated experimentally. Using ultrashort laser pulses with varying fluence, Bloch skyrmions, stabilised mainly by dipole-dipole interactions, were created in ferrimagnetic TbFeCo[34,35]. Skyrmions have also been experimentally created using laser heat pulses in ferromagnetic FeGe, where Bloch skyrmions are stabilised by the bulk DMI[36]. In another experiment on CoFeB/Ta multilayers, where Néel skyrmions are stabilised by interfacial DMI, illumination with single-shot ultrafast laser pulses resulted in formation of skyrmion clusters, with skyrmion density dependent on the laser pulse fluence[37]. Importantly, there was no difference observed in skyrmion formation between laser pulses with different helicities, and the skyrmion creation mechanism was attributed to heating.

Here we investigate in detail the dynamical skyrmion creation process following ultrafast demagnetisation in antiferromagnetic (AFM) and ferromagnetic (FM) thin films, and show transient DW skyrmions are formed after recovery of magnetisation order. Whilst deterministic switching of magnetisation is possible, we discuss the question if a single Néel skyrmion can be deterministically created using ultrafast laser pulses. We show that for both linearly and circularly polarised laser pulses, the possibility of a skyrmion with DW skyrmion pair self-annihilation is a barrier to deterministic creation of a single Néel skyrmion. Using thousands of skyrmion creation events we further analyse the statistical properties of skyrmion creation as a function of laser pulse properties, both in AFM and FM cases, showing the final states obey a Poisson counting distribution.

## Results

**Dynamical skyrmion creation process.**

Theoretically, skyrmion creation using ultrafast laser pulses was investigated for Néel skyrmions using vortex laser pulses, where skyrmions are created due to the effective magnetic field[38,39], whilst creation of skyrmions in antiferromagnetic insulators due to the inverse Faraday effect has also been studied[40]. On the other hand, creation of Bloch skyrmions and antiskyrmions has been investigated in chiral and dipolar magnets[41], where Langevin dynamics were taken into account during a rectangular heat pulse. Here we investigate the creation of Néel skyrmions both in AFM and FM materials, taking into account stochasticity both during and after a heat pulse. First we concentrate on the AFM case, then compare the skyrmion creation mechanism to that in the FM case.

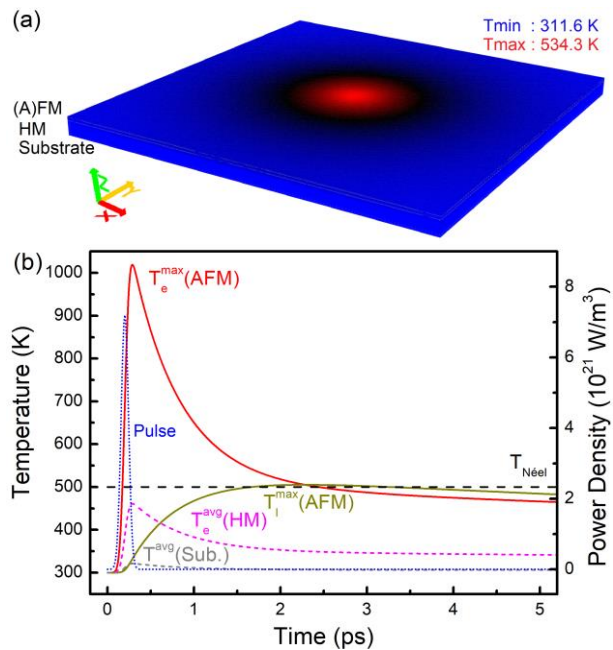

**Figure 1. Ultrafast Heat Pulse in a Magnetic System. (a)** Temperature during a Gaussian profile laser pulse in an (anti)ferromagnetic (2 nm) / heavy metal (8 nm) / substrate (40 nm) structure. **(b)** Typical ultrafast laser pulse and temperature time dependence in the layers: magnetic layer maximum electron and lattice temperatures, heavy metal layer average electron temperature and substrate average temperature.

The geometry studied is shown in Figure 1a. Here we use a 1 μm², 2 nm thick magnetic thin film on top of an 8 nm thick Pt layer on a SiO$_2$ substrate, 40 nm thick. Further details and material parameters are given in the Methods section. The heat transport is solved using the two-temperature model[42,43], with continuity of heat flux and temperature across the interfaces, and Robin boundary conditions on the exposed surfaces of the magnetic layer and substrate[44]. The electron temperature is used in a two-sublattice stochastic Landau-Lifshitz-Bloch model, which includes contributions from dipole-dipole interactions, direct and interfacial DMI exchange, antiferromagnetic exchange including both homogeneous and non-homogeneous inter-lattice contributions, uniaxial anisotropy, and applied field. Full details are given in the Methods section and Supplementary Information. The effect of a linearly polarised laser pulse is included as a Gaussian heat source in the electron temperature heat equation as:

$$S = P_0 \exp\left(\frac{-|\mathbf{r}-\mathbf{r}_0|}{d^2/4\ln(2)}\right) \exp\left(\frac{-(t-t_0)^2}{t_R^2/4\ln(2)}\right) \quad (W/m^3),  \qquad (2)$$

where $d$ and $t_R$ are full-width at half-maximum (FWHM) values. Initially we fix $d$ to 800 nm and $t_R$ to 100 fs – later we also consider narrower and longer pulses, and also circularly polarised laser pulses by inclusion of the associated magneto-optical field. Curves for the electron and lattice temperatures in the different layers are shown in Figure 1(b). Due to the small heat capacity of the electron bath, the maximum electron temperature rises very rapidly during the heat pulse, exceeding the Néel temperature for up to a few picoseconds, before converging towards the slower changing lattice temperature, and eventually back towards room temperature on a longer time scale lasting beyond 1 nanosecond.

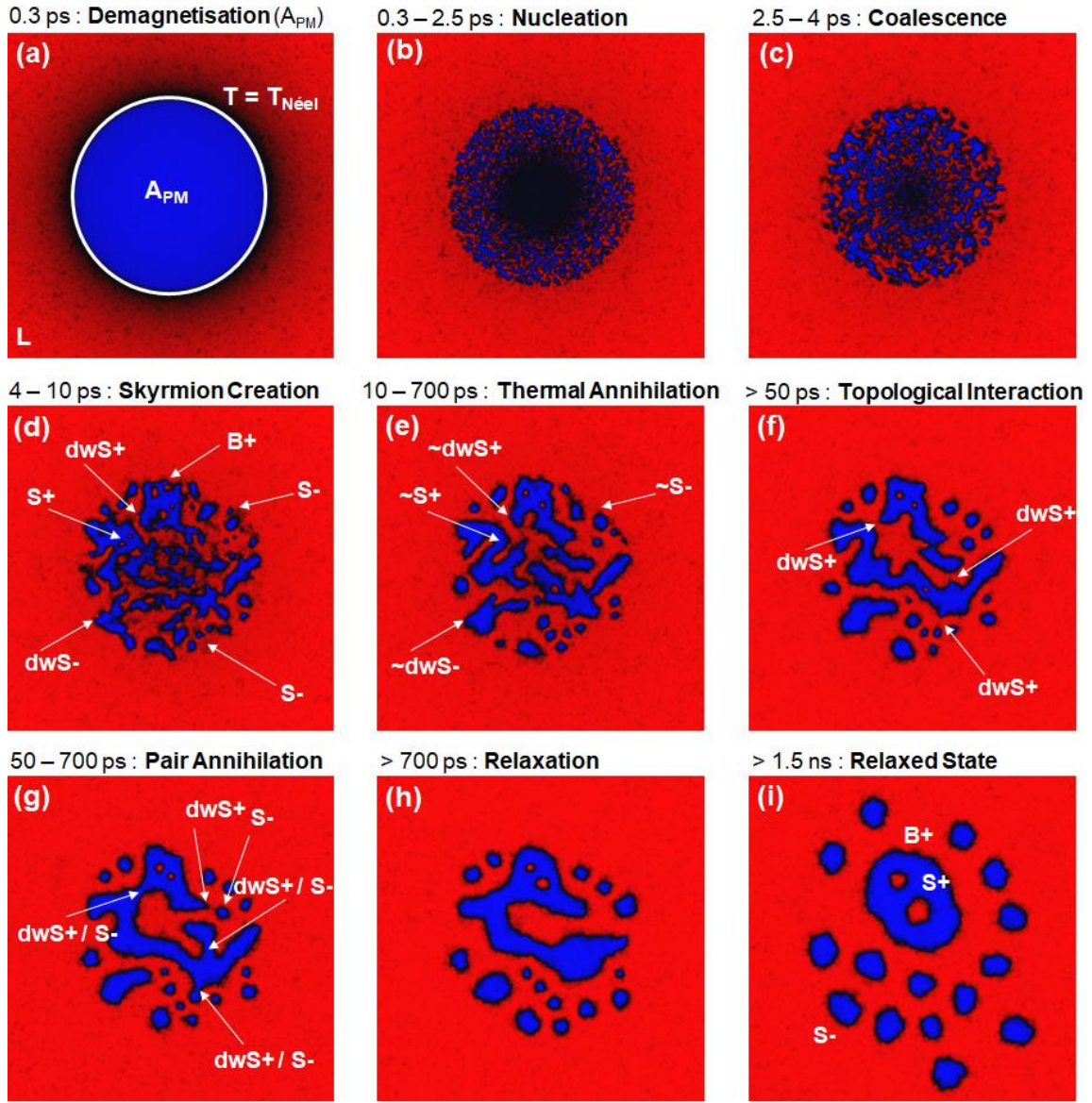

**Figure 2. Ultrafast Antiferromagnetic Skyrmions Creation.** Exemplification of the different stages observed after a linearly polarised (L) high power ($7\times10^{21}$ W/m$^3$) laser pulse, showing the z magnetisation component on sub-lattice A. **(a)** Demagnetisation, showing the paramagnetic area ($A_{PM}$). **(b)** Nucleation of reversed domains as the temperature drops below $T_{Néel}$. **(c)** Coalescence of nucleated domains. **(d)** Skyrmion creation due to DMI, showing examples of S±, dwS±, and B+. **(e)** Thermal annihilations of skyrmions. **(f)** Topological interactions (annihilation and repulsion) become important over a longer time scale. **(g)** Pair annihilations, showing dwS+ / S- annihilations. **(h)** Relaxation towards the final state starts after a stable topological charge is reached. **(i)** Relaxed state with minimal skyrmion distortions and larger spacings due to repulsive topological interactions reached on a time-scale of ns.

A typical skyrmion creation process is shown in Figure 2 for a single high power linearly polarised laser pulse ($7\times10^{21}$ W/m$^3$), plotting the z component of magnetisation on sub-lattice A; typically we distinguish several stages in the skyrmion creation process. The timing of these stages discussed below depends on the laser power, and even for the same power there is some variation and uncertainty in describing exactly where a stage starts and ends, although we can more precisely define them by analysing the time dependence of the topological charge. The first stage is demagnetisation during the applied laser pulse, where the temperature rises rapidly and phase transition to the paramagnetic state occurs for T > $T_{Néel}$; at the end of the laser pulse $A_{PM}$ is reached, which is the maximum area around the central spot with T > $T_{Néel}$. From this point on the temperature starts to decrease and, as it drops below $T_{Néel}$, areas with reversed magnetisation direction emerge, first at the outer boundary of $A_{PM}$, then moving inwards towards the center as the area with T > $T_{Néel}$ contracts. The nucleation and growth model, where reversed domains are nucleated through thermal activation over an energy barrier following an Arrhenius law[45,46], cannot be used to explain the nucleation stage observed here. Such processes occur on longer time-scales of 10 – 100 ps and longer, whilst the ultrafast nucleation stage occurs on a time-scale of ~1 ps. The nucleation stage is due to the much faster longitudinal relaxation process[47,48], and requires quenching of net magnetisation. As the temperature cools below the phase transition temperature, the nucleons of reversed magnetisation formed in the quenched state give rise to domains with reversed magnetisation. The next stage consists of coalescence of nucleated domains where, due to the high density of nucleation centres, the nucleated domains very rapidly coalesce into larger domains. Such localisation and coalescence processes have also been identified recently in ferrimagnetic alloys[49]. Since the temperature is not uniform, as for the nucleation stage, coalescence first starts at the outer boundaries of $A_{PM}$, proceeding towards the center as seen in Figure 2(c). As larger domains emerge, topologically protected structures become gradually distinguishable, formed from the disordered domain wall magnetisation under the effect of DMI. This is reflected by a rapid increase in the topological charge as seen in Figure 3(a) for the skyrmion creation stage.

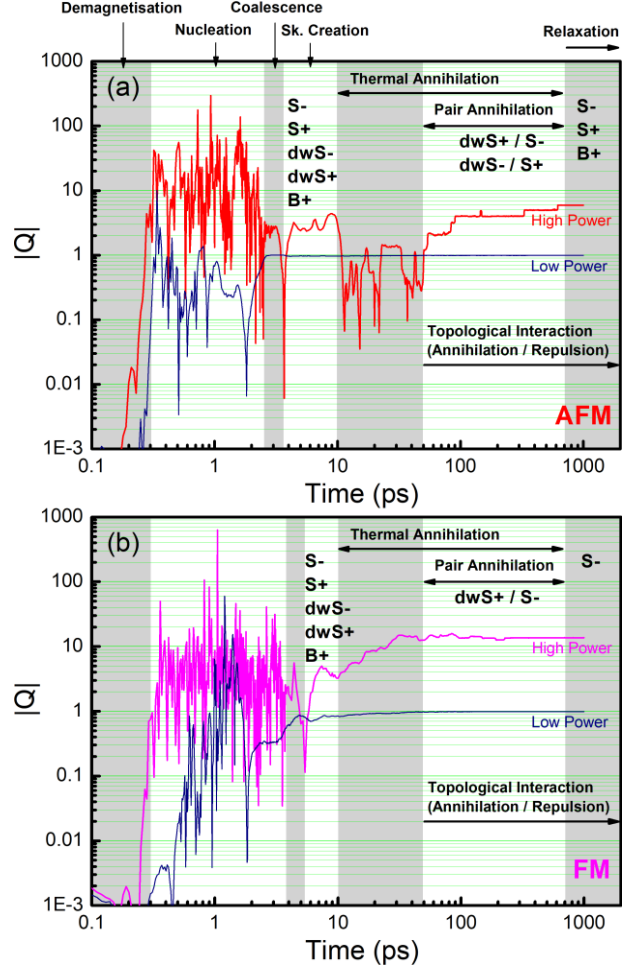

**Figure 3. Topological Charge Dynamics**. The different stages in the skyrmion creation process are reflected in the change in computed $Q$ values, here plotting $|Q|$ as a function of time for both **(a)** AFM sub-lattice A, and **(b)** FM at high ($7\times10^{21}$ W/m$^3$) and low ($3\times10^{21}$ W/m$^3$) laser powers with linear polarisation. During the nucleation process the $|Q|$ value changes randomly due to thermal fluctuations and highly disordered magnetisation, and is not a well-defined measure. The nucleated domains coalesce, typically resulting in a sharp drop in $|Q|$ towards the real topological charge at this stage: zero. Under the action of DMI, topological objects start to emerge as Néel domain walls are formed, resulting in a sharp increase in $|Q|$ which signals the start of the skyrmion creation stage. Many of the topological objects are gradually destroyed due to thermal activation, resulting in fluctuations in $|Q|$. On a longer time scale topological interactions become important, including pair annihilations and repulsive forces. Finally, a stable topological charge is reached, with relaxation continuing as a result of energy minimization, and in particular repulsive topological interactions. The same stages can be observed at low powers, but the time scales are shorter.

Here $|Q|$ is plotted as a function of time, computed using Equation (1), showing features for typical skyrmion creation processes at high ($7\times10^{21}$ W/m$^3$) and low ($3\times10^{21}$ W/m$^3$) laser powers. During the nucleation stage the *Q* value is ill-defined due to the high degree of magnetisation disorder. As the coalescence stage starts, the noise in the computed *Q* value very rapidly settles, and a sharp drop towards zero is observed in the vast majority of cases, both for AFM and FM, and both at high and low powers. After this drop, typically a sharp rise towards a well-defined value of |*Q*| signifies the start of the skyrmion creation stage as noted above. At the end of the skyrmion creation process the reversed domains are separated by Néel domain walls, although very significant distortions are still present in the skyrmions at this stage as seen in Figure 2(d). During the skyrmion creation stage, for both AFM and FM cases, the most common type of skyrmion formed is the S- skyrmion – for a full list see Table 1. S+ skyrmions are also formed, although these are much rarer since they require nucleation of a domain within another reversed domain. When they are created however, they form skyrmion bags, B+. It's also possible for a B- skyrmion bag to be formed, however this is extremely rare since they require an S+ within an S-, within an S+ skyrmion – in three thousand events this was observed only twice, and in both cases a B-(0) skyrmionium was formed.

**Table 1. List of Observed Skyrmions.** The skyrmions are characterised on sub-lattice A for the AFM case, showing the topological charge *Q(A)*, and indicating the topological objects found in the relaxed state for AFM and FM cases.

| *Symbol* | *Name* | *Q(A)* | **Relaxed State** |
|---|---|---|---|
| **S-** | Skyrmion | **-1** | AFM, FM |
| **S+** | Skyrmion | **+1** | AFM |
| **dwS-** | Domain Wall Skyrmion | **-1** | None |
| **dwS+** | Domain Wall Skyrmion | **+1** | None |
| **B-(|Q|)** | Skyrmion Bag (Skyrmionium) | **≤0 (0)** | AFM (probability < $10^{-3}$) |
| **B+(|Q|)** | Skyrmion Bag (Skyrmionium) | **≥0 (0)** | AFM |

**Domain wall skyrmion.**

Another type of topological object formed, which is ubiquitous in both AFM and FM cases, is the DW skyrmion[20-23] with topological charge of ±1, i.e. dwS±; examples are indicated in Figure 2(d). The domain wall skyrmion is a 360º degree rotation of the in-plane magnetisation components transverse to an out-of-plane Néel domain wall, stabilised by the DMI and topologically protected. Figure 4 shows examples of dwS+ relaxed at 0 K, attached to an antiferromagnetic skyrmionium in Figure 4(a), and to a simple ferromagnetic Néel domain wall in Figure 4(b). Whilst a large population of dwS± are created during the skyrmion creation stage, these objects never survive in the relaxed state owing to a combination of rapid thermally activated collapse, as well as pair annihilations with S±, which could explain why such topological structures have not yet been observed in experiments. Skyrmions and DW skyrmions with opposite topological charge experience an attractive topological interaction, mainly owing to spatial gradients in the DMI energy, whilst dipole-dipole interactions also play a role in the FM case at larger separations[50]. When sufficiently close this results in a pair annihilation with no net change in the topological charge. Examples of such annihilations, in particular dwS+ / S- annihilations, have been identified in Figure 2(f),(g) – as result of a pair annihilation, the isolated S- skyrmion is effectively absorbed by the S- skyrmion with the dwS+ attached, which results in the unwinding of the dwS+ structure. On the other hand, skyrmions and DW skyrmions with the same topological charge experience a repulsive topological interaction, similar to that observed between skyrmions. Thus in Figure 4(a) the B+(0) – dwS+ pair is stable owing to the topological repulsion between the dwS+ and the S+ skyrmion inside the skyrmionium. If on the other hand a single dwS+ is attached to an S- skyrmion, the pair is not stable even at 0 K – owing to the topological attraction such a pair rapidly contracts, resulting in self-annihilation where both the S- and dwS+ are absorbed into the background state. However, if multiple dwS+ objects are attached to a single S-, the structure becomes stable at 0 K due to the repulsive interaction between the dwS+ objects, which prevents collapse of the S- skyrmion.

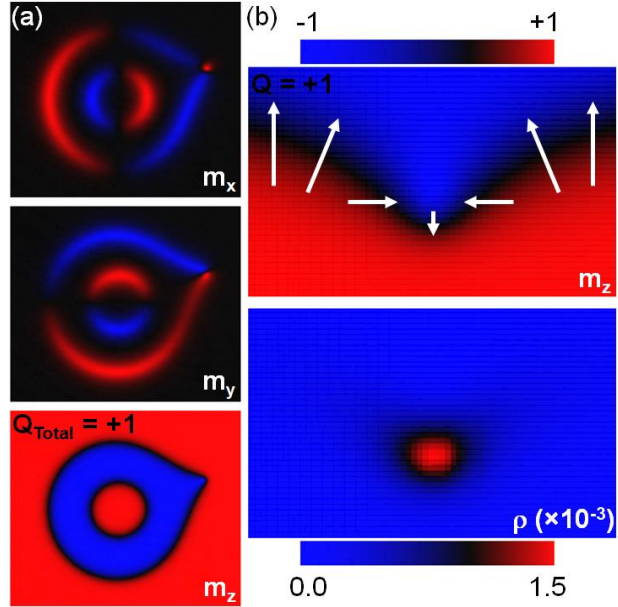

**Figure 4. Domain Wall Skyrmion. (a)** Stable antiferromagnetic skyrmionium – DW skyrmion pair (B+(0) – dwS+) with total topological charge of +1 on sub-lattice A, showing the individual magnetisation vector components. **(b)** Ferromagnetic DW skyrmion showing the z component, with overlaid sketch of the in-plane components of magnetisation. The panel below shows the corresponding topological charge density, ρ. All magnetisation configurations are relaxed at T = 0 K.

**Thermal decay.**

After creation, many topological objects are thermally annihilated as exemplified in Figure 2(e), including skyrmions far away from equilibrium, whilst dwS± objects are inherently susceptible to rapid thermally activated collapse as we discuss below. This results in fluctuations in the computed value of |Q| as shown in Figure 3, both for AFM and FM: due to a preponderance of S-, collapse of dwS+ and S+ increases the |Q| value, whilst collapse of dwS- and S- decreases it. Finally, as all the pair and thermal annihilations of unstable objects have completed, the final $Q$ value is reached; from this point the skyrmions continue to relax – Figure 2(h) – expanding beyond $A_{PM}$ under the action of repulsive topological interactions, eventually reaching a relaxed state on a time-scale of nanoseconds. It has been shown that skyrmions decay through thermal activation as described by an Arrhenius law, both for Néel FM skyrmions[51,52], Bloch FM skyrmions[53] and AFM skyrmions[54], with experimental verification available in a chiral magnet[55]. Here we show DW skyrmions also decay through thermal activation following an Arrhenius law. Figure 5 shows the thermal decay of AFM dwS+, where a skyrmionium with 6 attached dwS+ is isolated, and a typical thermal decay is shown in Figure 5(a) at 325 K.

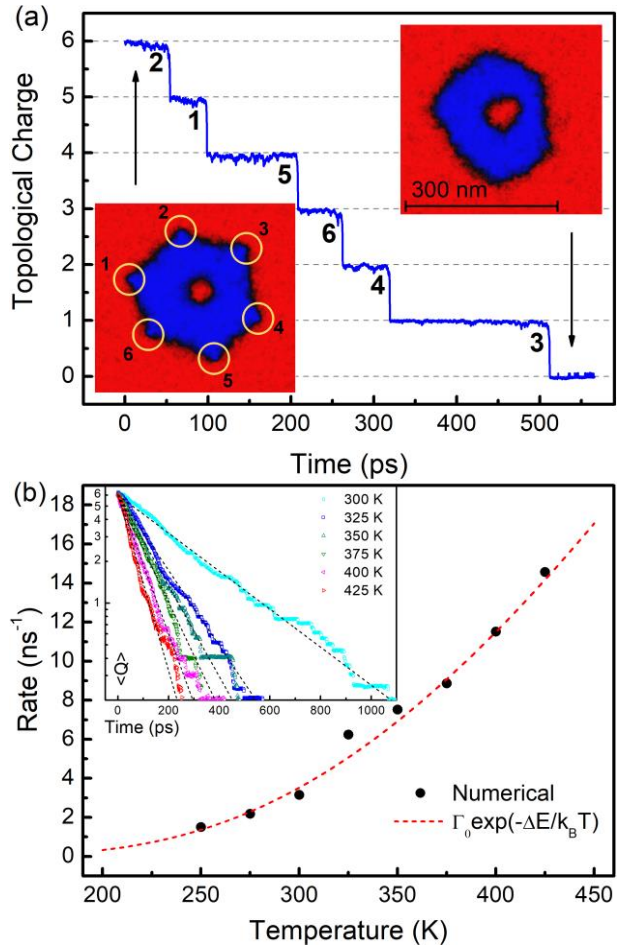

**Figure 5. Thermal Decay of Domain Wall Skyrmions.** Thermal decay of an AFM skyrmionium – DW skyrmion complex with total topological charge of +6 is shown, for **(a)** single decay at 325 K, and **(b)** averaged decay for a range of temperatures. The decay is described by an exponential process with a decay rate constant following an Arrhenius law with energy barrier $\Delta E = 2\times10^{-20}$ J and attempt frequency $\Gamma_0 = 4\times10^{11}$ s$^{-1}$.

Whilst the DW skyrmions are also topologically protected, they can collapse through thermal activation at a faster rate compared to FM or AFM skyrmions, both of which have life-times on time-scales of ns or longer at room temperature, depending on material parameters. The thermally activated collapse of skyrmions arises through a gradual contraction in the diameter, until flipping of the core magnetisation results in loss of the topological structure[51]; another possibility is by nucleation of a singularity resembling a hedgehog Bloch point[52]. On the other hand, the thermal collapse of a DW skyrmion is much simpler since the topological charge density is highly localised, as seen in Figure 4(b). Thus we only require the central in-plane components to flip, which is driven by thermal activation over an energy barrier. To see this, we compute the average topological charge over 30 decays of the skyrmionium – DW skyrmion complex for each temperature, with results shown in Figure 5(b). The thermal decay is exponential with a single decay rate dependent on temperature, confirming the thermal collapse of the 6 dwS+ is independent of the number of dwS+ attached to the skyrmionium. The decay rate follows an Arrhenius law as a function of temperature, $\Gamma = \Gamma_0 \exp(-\Delta E/k_B T)$, where we obtain the attempt frequency $\Gamma_0 = 4\times10^{11}$ s$^{-1}$ and the energy barrier $\Delta E = 2\times10^{-20}$ J. This energy barrier is comparable to that obtained for FM skyrmions[51], however the attempt frequency is much higher resulting in rapid thermal decay on time scales of ps at room temperature. It should be noted dwS- are significantly rarer than dwS+, because in a dwS- – S- pair the added repulsive topological interaction, which favours flipping of the dwS center, causes the dwS- object to collapse much quicker than a dwS+.

**Single Néel skyrmion creation.**

Whilst a large population of skyrmions is typically created at high powers, it is possible to create a single skyrmion at low powers, as shown in Figure 3. It is important to note however that this is not a deterministic process in the cases studied here, both for AFM and FM: once a skyrmion is created there is a distinct probability of thermally activated collapse after creation when the skyrmion is still far away from equilibrium, which occurs on time scales of up to 100 ps. This is in contrast to a previous work on Bloch skyrmions[41] which investigated conditions under which a single skyrmion can be created with 100% efficiency as a result of an ultrafast heat pulse. There the temperature was set to zero after the heat pulse, thus removing the thermal activation process which can lead to skyrmion collapse for the cases we've studied. As shown in Figure 1 the temperature decays on a much longer time scale compared to the ultrafast heat pulse, with the temperature changing slowly after the initial temperature spike. The stochasticity of magnetisation dynamics must be taken into account during this stage, which can result in a range of possible outcomes for the final relaxed state. Experimental results on Néel skyrmion formation after an ultrafast heat pulse have also shown random skyrmion nucleation at low laser fluences[37], however the formation of Bloch skyrmion bubbles with 100% efficiency was reported in experiments on a dipolar magnet[34]. Another possibility, intrinsic to the Néel skyrmions studied here, is for a S- – dwS+ pair to be nucleated which results in collapse of the skyrmion and finally self-annihilation even in the absence of stochasticity.

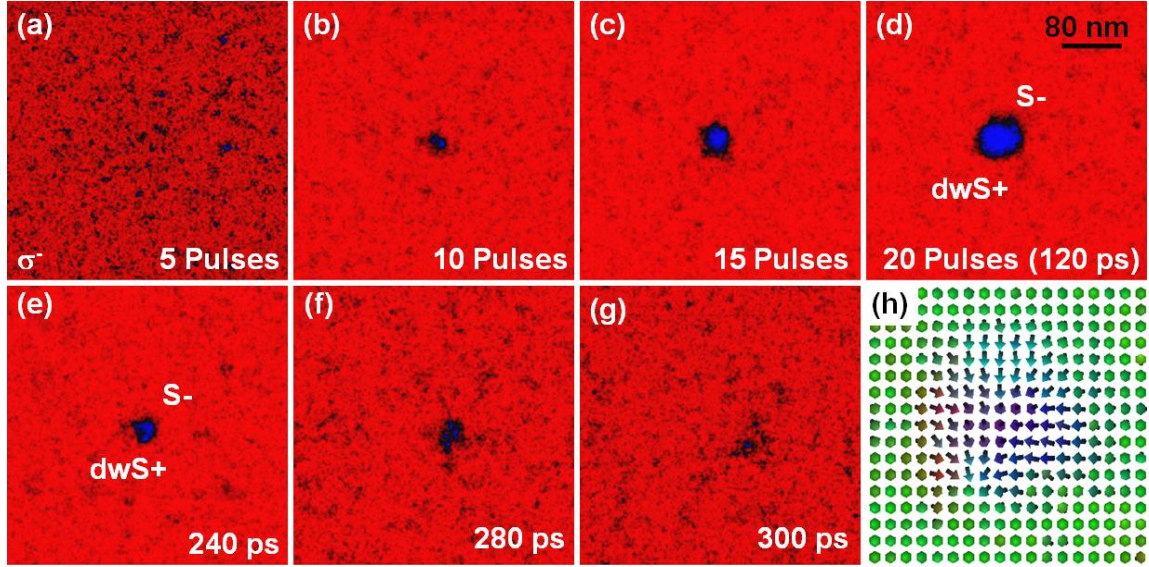

**Figure 6. Skyrmion Self-Annihilation with a Domain Wall Skyrmion.** Creation of a ferromagnetic Néel skyrmion using a train of 20 circularly polarised laser pulses, heating the material close to, but below the Curie temperature. A negative helicity ($\sigma^-$) is used which results in deterministic switching of magnetisation with sufficient number of pulses. **(a)-(d)** Magnetisation state, showing the z component, after the indicated number of pulses. After the last pulse an S- – dwS+ pair is visible, with a close-up magnetisation configuration shown in **(h)**. **(e)-(g)** Collapse and self-annihilation of S- – dwS+ pair due to topological attraction.

The remaining question is whether a single Néel skyrmion can be created deterministically using a train of laser pulses, instead of a single shot pulse, and additionally using circularly polarised light where HD-AOS arises. It has been shown experimentally that no helicity dependence exists in a FM material with interfacial DMI[37] for single-shot ultrafast laser pulses. If the main mechanism giving rise to a helicity dependence is the inverse Faraday effect, where a perpendicular magneto-optical field is present during the laser pulse, this observation is not surprising since precessional magnetisation processes require significantly longer time to respond to the magnetic field than available in a single sub-ps laser pulse. Instead, the accumulated effect of a train of laser pulses is required to deterministically switch magnetisation with a helicity dependence in FM materials[56-58]. Thus, whilst a large area can be switched deterministically using a train of laser pulses[29], it remains an open question whether a single Néel skyrmion can be deterministically created under appropriate conditions. It has already been shown the HD-AOS reversed domain size needs to be larger than the laser spot size[59]. Here we show that even with a train of circularly polarised laser pulses which heat the material close to the phase transition temperature, the possibility of nucleating a S- – dwS+ pair or complex cannot be disentangled from the possibility of creating just one S-. Due to the circular polarisation of the laser pulse a strong perpendicular magneto-optical field is present, given by $H_{MO} = \sigma^{\pm} H_{MO}^{0} f_{MO}(\mathbf{r},t)\hat{\mathbf{z}}$ [25], where $f_{MO}$ has the spatial and temporal dependence given in Equation (2), and $\sigma^{\pm} = \pm 1$. The results in Figure 6 are shown for the FM case, where we apply a sequence of 20 circularly polarised laser pulses at 6 ps intervals with negative helicity ($\sigma^-$) and strength of 10 MA/m. An out-of-plane bias field of 100 kA/m is used, and we have checked the positive helicity does not result in switching of magnetisation. Here $t_R = 500$ fs, and $d = 100$ nm, which is slightly larger than the ideal skyrmion diameter of 80 nm. Throughout the pulse sequence in Figure 6 the temperature doesn't exceed the Curie temperature. The cumulative effect of laser pulses is to gradually reduce the magnetisation in the central spot due to the higher temperature, and eventually a small reversed domain is nucleated under the strong magneto-optical field – Figure 6(b). As further pulses are applied this reversed domain grows until a maximum size is reached – Figure 6(c),(d). Whilst this results in a single S- skyrmion in some cases, due to the large degree of disorder at the skyrmion boundary there is a distinct probability of nucleating one or more dwS, as is the case in Figure 6(d) – a close-up of the magnetisation structure is shown in Figure 6(h), clearly identifying a S- – dwS+ pair. As a result of the topological attraction between the S- and dwS+, the skyrmion quickly collapses – Figure 6(e),(f) – and is annihilated – Figure 6(g). Finally, there is the possibility that creating a large enough skyrmion avoids a self-annihilation collapse if the DW skyrmion life-time is significantly shorter than the skyrmion collapse time. We have also investigated this, however typically for a large skyrmion the border tends to have significant distortions after ultrafast laser pulses, with many dwS± present, which rather than cause a complete collapse typically result in the skyrmion splitting into multiple skyrmions; material defects and multi-layers will further complicate this process, however this is beyond the scope of this work.

**Statistical properties.**

We've already remarked on the differences between skyrmion creation at high and low laser powers. Here we systematically study the statistical properties of skyrmion creation as a function of heat pulse properties, first for the AFM case, and in the next section we investigate the differences for the FM case. For this, a laser pulse with linear polarisation is applied, and when a stable topological charge is reached the number of skyrmions created is computed. For each heat pulse setting this process is repeated up to 50 times.

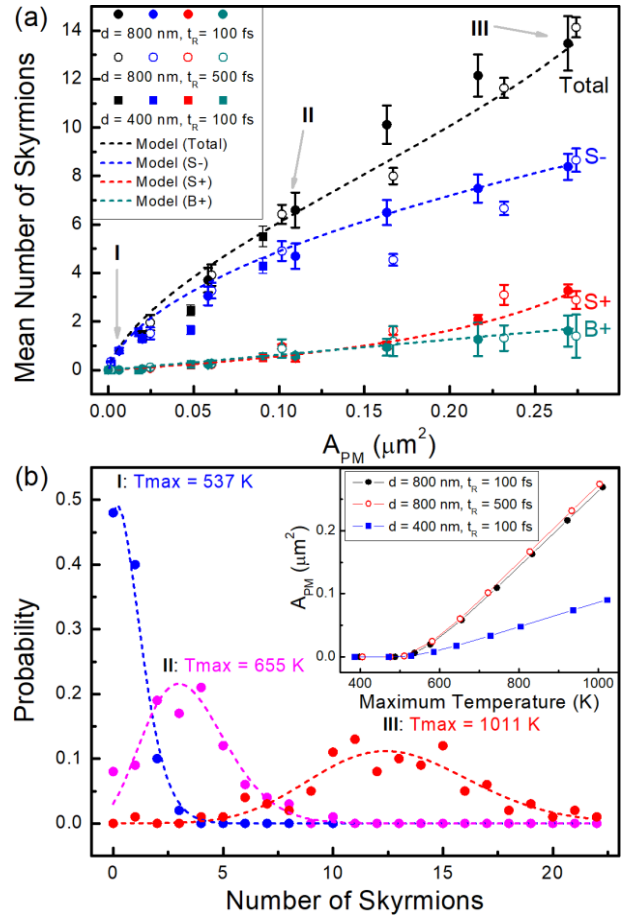

**Figure 7. Antiferromagnetic Skyrmions Statistics. (a)** Mean number of skyrmions created is shown as a function of $A_{PM}$ for FWHM pulse values of $d$ = 400 nm, 800 nm, and $t_R$ = 100 fs, 500 fs. The different types of stable topological objects formed are shown, namely S-, S+, and B+. The dashed lines are obtained from the model in Equation (3). **(b)** Skyrmion creation probability distributions are shown for 3 selected pulse strengths, with fitted Poisson distributions – the values in (a) are the fitted mean rates together with fitting uncertainties, obtained from 50 repetitions for each pulse setting. The inset shows $A_{PM}$ as a function of maximum temperature reached for the different pulses used.

The results are shown in Figure 7. Here we use 2 values of pulse width and duration, namely $d$ = 400 nm, 800 nm, and $t_R$ = 100 fs, 500 fs, and vary the pulse power. The variation of $A_{PM}$ with maximum temperature reached for the different pulse settings is shown in the inset to Figure 7(b), and Figure 7(a) shows the mean number of skyrmions created as a function of $A_{PM}$. Whilst there is a significant statistical uncertainty in obtaining the mean skyrmion creation rates due to the limited number of events included, the rates for S-, S+, and B+ follow the same trend lines respectively when plotted as a function of $A_{PM}$, even though the pulse characteristics are otherwise very different. This can be understood from the dynamical skyrmion creation process, in particular the nucleation stage – see Figure 3(b). Since the average number of nucleated domains is dependent on $A_{PM}$, then so is the final number of stable skyrmions. The maximum temperature reached doesn't have a noticeable effect on the skyrmion creation rates for the same $A_{PM}$, serving only to reach a quenched magnetisation state. Another important observation, is the number of skyrmions created follows a Poisson distribution, i.e. $P(n) = \lambda^n e^{-\lambda}/n!$, where n is the number of skyrmions created and $\lambda$ is the mean skyrmion creation rate. Examples are shown in Figure 7(b) for 3 different pulse powers – the mean skyrmion creation rate is obtained from a Poisson distribution fit.

$$N_{S\pm}^{AFM} = \frac{A_{PM}}{A_{S\pm}^{AFM}} \left[ 1 + r_\pm \left( N_{S\pm}^{AFM} - 1 \right) \right]^{\pm 1} \tag{3}$$

Next, we develop a simple phenomenological model to describe the variation in mean number of S-, S+ and B+ objects created as a function $A_{PM}$. As $A_{PM}$ increases, the number of more complex topological objects created, B+, increases, which mirrors the experimental observations for Bloch skyrmions[34]. The increase in mean number of S- skyrmions is not linear however, but slows down as $A_{PM}$ increases, i.e. the density of S- decreases with $A_{PM}$. As the number of S- skyrmions increases, due to the probability of nucleating S+ skyrmions inside them, the number of S+ also increases. The resultant B+ objects tend to occupy a significantly larger area, which reduces the available area for S- skyrmions, and thus decreasing their density. On the other hand, the density of S+ skyrmions increases slightly with $A_{PM}$ due to the topological pressure experienced inside a skyrmion bag. The B+ objects are constrained by topological repulsion from surrounding S- skyrmions. In turn this results in increased density of the contained S+ skyrmions, as they experience a topological repulsion from neighbouring S+, as well as from the Néel border of the containing B+ – for example see Figure 2(d). The model is shown in Equation (3). Here $A_{S\pm}^{AFM}$ are the paramagnetic areas required to create one S± skyrmion respectively on average, whilst $r_+ \cong$ 0.3 and $r_- \cong 0.4$ are fitting factors representing the increase, respectively decrease, in S± skyrmion density with increasing $A_{PM}$. The increase in B+ with $A_{PM}$ is assumed to be approximately linear. We've also repeated these calculations for varying DMI strength, with results shown in the Supplementary Information, where the same *r* fitting factors are used. A reasonable agreement is obtained between the numerical results with statistical information extracted from 50 events for each set of simulation parameters, and the simple phenomenological model in Equation (3).

**Comparison with ferromagnetic skyrmions.**

In the case of FM, in addition to DMI, skyrmions also interact through the dipole-dipole interaction, which is an additional source of topological repulsion between skyrmions of same topological charge. Notwithstanding, there are strong similarities to the AFM case, some of which have been discussed in relation to Figure 3. In terms of dynamical processes, the same stages seen in Figure 2 are also observed for the FM case. The most important difference however, in the relaxed state only S- skyrmions are observed. Typical resultant relaxed stages are shown in Figure 8(b),(c): for the FM case the skyrmions relax into a hexagonal lattice as observed experimentally[37], whilst for the AFM case the final state is composed of a combination of S-, S+, and B+ objects. As for AFM, dwS objects exhibit a quick thermal decay, and also participate in skyrmion – DW skyrmion pair annihilations. S+ skyrmions on the other hand are not stable, since the direction of the applied magnetic field favours stabilisation of S- skyrmions only, and collapse on a time-scale shorter than 1 ns.

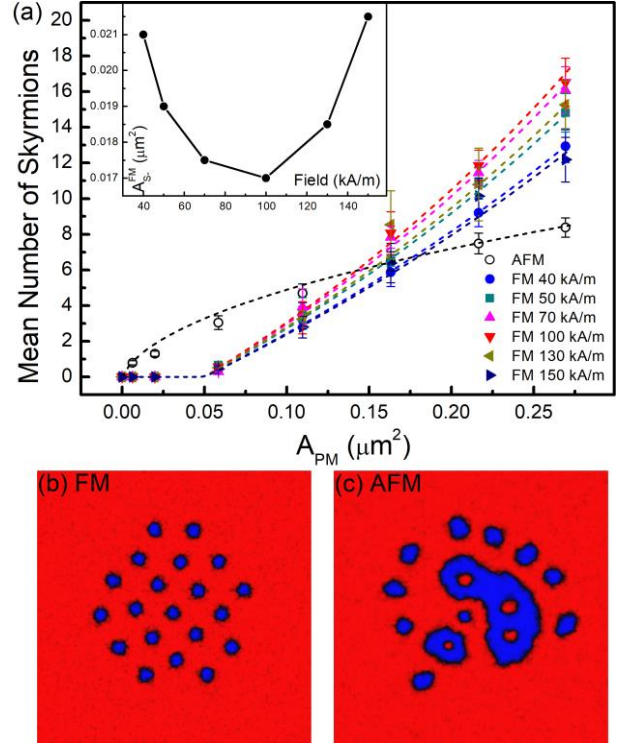

**Figure 8. Ferromagnetic and Antiferromagnetic Skyrmions Comparison. (a)** Mean number of S- skyrmions as a function of $A_{PM}$, for out-of-plane field strength varying from 40 kA/m up to 150 kA/m. The dashed lines are obtained from the model in Equation (**4**). For comparison the S- antiferromagnetic skyrmion mean rates for D = 1 mJ/m$^2$ are also shown. The inset shows the area required to obtain one skyrmion on average as a function of applied field. **(b)** Typical created FM skyrmion cluster, and **(c)** typical created AFM skyrmion collection at high power (7×10$^{21}$ W/m$^3$).

In terms of statistical properties, the Poisson counting distribution is also obeyed by the resultant FM skyrmion states. The mean number of S- skyrmions is plotted in Figure 8(a) as a function of $A_{PM}$, and compared to the AFM case. In contrast to the AFM case, a threshold paramagnetic area $A_0 \cong 0.05$ μm$^2$ is required for any skyrmions to be formed. This is due to the paramagnetic susceptibility resulting in a net magnetic moment when a field is applied, and therefore a larger maximum temperature is required to reach the quenched magnetic state necessary for nucleation of reversed domains. We can recover the same physical picture by redefining the temperature value from which $A_{PM}$ is calculated, or alternatively we can shift the zero point to $A_0$. As we've verified, the value of $A_0$ saturates quickly with fields greater than 1 kA/m, however the calculation of $A_0$ dependence on field and maximum temperature is beyond the scope of this study and is left for future work. As $A_{PM}$ increases, the mean number of S- skyrmions is found to increase slightly faster than linear, which is a result of increasing skyrmion density with larger $A_{PM}$. This is also in agreement with experimental observations[37], where long distance dipole-dipole interactions result in compression of skyrmions at the center of $A_{PM}$. Similar to the AFM case, the applicable phenomenological model is shown in Equation (4), plotted in Figure 8.

$$N_{S-}^{FM} = \frac{(A_{PM} - A_0)}{(A_{S-}^{FM} - A_0)}[1 + r(N_{S-}^{FM} - 1)] \qquad (4)$$

Here a much smaller $r$ factor of 0.02 is obtained, since the increase in density due to dipole-dipole interactions occurs over much larger distances compared to the DMI origin in the AFM case. The area required for single skyrmion creation, $A_{S-}^{FM}$, is dependent on the applied field strength as shown in the inset to Figure 8(a). As the applied field increases, the skyrmion diameter decreases, which results in a decreased $A_{S-}^{FM}$ up to 100 kA/m. However, at the same time the smaller skyrmions are more susceptible to thermally activated collapse, thus further increase in the applied magnetic field requires increasingly larger $A_{S-}^{FM}$ to maintain the same mean skyrmion creation rate.

## Methods

**Model.**

We use a two-sublattice stochastic Landau-Lifshitz-Bloch (sLLB) equation, based on the LLB equation from Refs.[60,61], applicable for antiferromagnetic, ferrimagnetic, as well as binary ferromagnetic alloys. We include both homogeneous and non-homogeneous inter-lattice exchange contributions, and recast the model in terms of accessible micromagnetic parameters, above and below the phase transition temperature. The explicit 2-sublattice sLLB equation is given in Equation (5) in terms of the macroscopic magnetisation, where we denote the 2 sublattices as $i = A, B$.

$$\frac{\partial \mathbf{M}_i}{\partial t} = -\tilde{\gamma}_i \mathbf{M}_i \times \mathbf{H}_{eff,i} - \tilde{\gamma}_i \frac{\tilde{\alpha}_{\perp,i}}{M_i} \mathbf{M}_i \times \left( \mathbf{M}_i \times \left( \mathbf{H}_{eff,i} + \mathbf{H}_{th,i} \right) \right)$$
$$+ \gamma_i \frac{\tilde{\alpha}_{\parallel,i}}{M_i} \left( \mathbf{M}_i \cdot \mathbf{H}_{\parallel,i} \right) \mathbf{M}_i + \boldsymbol{\eta}_{th,i} \quad (i = A, B) \tag{5}$$

The reduced gyromagnetic ratio is given by $\tilde{\gamma}_i = \gamma_i / (1 + \alpha_{\perp,i}^2)$, and the reduced transverse and longitudinal damping parameters by $\tilde{\alpha}_{\perp(\parallel),i} = \alpha_{\perp(\parallel),i} / m_i$, where $m_i(T) = M_i(T) / M_{S,i}^0$, with $M_{S,i}^0$ denoting the zero-temperature saturation magnetisation, and $M_i \equiv |\mathbf{M}_i|$. The damping parameters are continuous at $T_N$ – the phase transition temperature – and given by:

$$\alpha_{\perp,i} = \alpha_i \left( 1 - \frac{T}{3(\tau_i + \tau_{ij} m_{e,j} / m_{e,i}) \tilde{T}_N} \right), \quad T < T_N$$
$$\alpha_{\parallel,i} = \alpha_i \left( \frac{2T}{3(\tau_i + \tau_{ij} m_{e,j} / m_{e,i}) \tilde{T}_N} \right), \quad T < T_N \tag{6}$$
$$\alpha_{\perp,i} = \alpha_{\parallel,i} = \frac{2T}{3T_N}, \quad T \geq T_N$$

We denote $\tilde{T}_N$ the re-normalized transition temperature, given by:

$$\tilde{T}_N = \frac{2T_N}{\tau_A + \tau_B + \sqrt{(\tau_A - \tau_B)^2 + 4\tau_{AB}\tau_{BA}}} \tag{7}$$

The micromagnetic parameters $\tau_i$ and $\tau_{ij} \in [0, 1]$, are coupling parameters between exchange constants and the phase transition temperature, such that $\tau_A + \tau_B = 1$ and $|J| = 3\tau k_B T_N$. Here $J$ is the exchange constant for intra-lattice ($i = A,B$) and inter-lattice ($i,j = A,B, i \neq j$) coupling respectively. For a simple antiferromagnet we have $\tau_A = \tau_B = \tau_{AB} = \tau_{BA} = 0.5$. The normalised equilibrium magnetisation functions $m_{e,i}$ are obtained from the Curie-Weiss law as:

$$m_{e,i} = B\left[\left(m_{e,i}\tau_i + m_{e,j}\tau_{ij}\right)3\widetilde{T}_N/T + \mu_i\mu_0 H_{ext}/k_B T\right], \tag{8}$$

where $B(x) = \coth(x) - 1/x$, and $\mu_i$ is the atomic magnetic moment. The magnetisation length is not constant, and can differ from the equilibrium magnetisation length, giving rise to a longitudinal relaxation field which includes both intra-lattice and inter-lattice contributions:

$$\mathbf{H}_{\parallel,i} = \left\{\frac{1}{2\mu_0\widetilde{\chi}_{\parallel,i}}\left(1 - \frac{m_i^2}{m_{e,i}^2}\right) + \frac{3\tau_{ij}k_B T_N}{2\mu_0\mu_i}\left[\frac{\widetilde{\chi}_{\parallel,j}}{\widetilde{\chi}_{\parallel,i}}\left(1 - \frac{m_i^2}{m_{e,i}^2}\right) - \frac{m_{e,j}}{m_{e,i}}\left(\hat{\mathbf{m}}_i.\hat{\mathbf{m}}_j\right)\left(1 - \frac{m_j^2}{m_{e,j}^2}\right)\right]\right\}\mathbf{m}_i, \quad T < T_N$$

$$\mathbf{H}_{\parallel,i} = -\left\{\frac{1}{\mu_0\widetilde{\chi}_{\parallel,i}} + \frac{3\tau_{ij}k_B T_N}{\mu_0\mu_i}\left[\frac{\widetilde{\chi}_{\parallel,j}}{\widetilde{\chi}_{\parallel,i}} - \frac{m_{e,j}}{m_{e,i}}\left(\hat{\mathbf{m}}_i.\hat{\mathbf{m}}_j\right)\right]\right\}\mathbf{m}_i, \quad T > T_N \tag{9}$$

Here $\hat{\mathbf{m}}_i = \mathbf{m}_i/m_i$, and the relative longitudinal susceptibility is $\widetilde{\chi}_{\parallel,i} = \chi_{\parallel,i}/\mu_0 M_{S,i}^0$, where:

$$k_B T \chi_{\parallel,i} = \frac{\mu_i B_i'\left(1 - 3\tau_j\widetilde{T}_N B_j'/T\right) + \mu_j 3\tau_{ij}\widetilde{T}_N B_i' B_j'/T}{\left(1 - 3\tau_i\widetilde{T}_N B_i'/T\right)\left(1 - 3\tau_j\widetilde{T}_N B_j'/T\right) - \tau_{ij}\tau_{ji}B_i'B_j'\left(3\widetilde{T}_N/T\right)^2}, \tag{10}$$

and $B_i' \equiv B_{m_{e,i}}'\left[\left(m_{e,i}\tau_i + m_{e,j}\tau_{ij}\right)3\widetilde{T}_N/T\right]$.

The effective field in Equation (5) is a sum of all the interaction fields, and given by $\mathbf{H}_{eff,i} = \mathbf{H}_{ext,i} + \mathbf{H}_{demag,i} + \mathbf{H}_{ani,i} + \mathbf{H}_{ex,i}$. In particular $\mathbf{H}_{demag,i}$ is the demagnetising field calculated for the net magnetization $(\mathbf{M}_A + \mathbf{M}_B)/2$, and applied equally to both sub-lattices. The uniaxial anisotropy field is given by:

$$\mathbf{H}_{ani,i} = \frac{2K_{1,i}}{\mu_0 M_{e,i}^2}\left(\mathbf{M}_i.\mathbf{e}_A\right)\mathbf{e}_A \tag{11}$$

Here $\mathbf{e}_A$ is the symmetry axis, $M_{e,i} = m_{e,i}M_{S,i}^0$, and $K_{1,i}$ follows the temperature dependence $K_{1,i} = K_{1,i}^0 m_{e,i}^3$. The exchange field includes both the isotropic direct exchange term, as well as the interfacial DMI exchange term. The direct exchange term includes the usual intra-lattice contribution, as well as homogeneous and non-homogeneous inter-lattice contributions, and is given by:

$$\mathbf{H}_{ex,i} = \frac{2A_i}{\mu_0 M_{e,i}^2}\nabla^2\mathbf{M}_i + \frac{4A_{h,i}}{\mu_0 M_{e,i}M_{e,j}}\mathbf{M}_j + \frac{A_{nh,i}}{\mu_0 M_{e,i}M_{e,j}}\nabla^2\mathbf{M}_j \tag{12}$$

The intra-lattice exchange stiffness $A_i$ has the temperature dependence $A_i = A_i^0 m_{e,i}^2$, whilst the inter-lattice exchange stiffnesses have the temperature dependences $A_{h(nh),i} = A_{h(nh),i}^0 m_{e,i} m_{e,j}$. The interfacial DMI exchange field is given by, applicable for systems with $C_{nv}$ symmetry:

$$\mathbf{H}_{iDMI,i} = -\frac{2D_i}{\mu_0 M_{e,i}^2}\left(\frac{\partial M_z}{\partial x}, \frac{\partial M_z}{\partial y}, -\frac{\partial M_x}{\partial x} - \frac{\partial M_y}{\partial y}\right), \quad (13)$$

with the DMI exchange parameter having the temperature dependence $D_i = D_i^0 m_{e,i}^2$.

Finally, the terms $\mathbf{H}_{th,i}$ and $\mathbf{\eta}_{th,i}$ are stochastic quantities with zero spatial, vector components, and inter-lattice correlations, and whose components follow Gaussian distributions with zero mean and standard deviations given respectively by:

$$H_{th,i}^{std.} = \frac{1}{\alpha_{\perp,i}}\sqrt{\frac{2k_B T(\alpha_{\perp,i} - \alpha_{\parallel,i})}{\gamma_i \mu_0 M_{S,i}^0 V \Delta t}}$$
$$\eta_{th,i}^{std.} = \sqrt{\frac{2k_B T \alpha_{\parallel,i} \gamma_i M_{S,i}^0}{\mu_0 V \Delta t}} \quad (14)$$

Here $V$ is the stochastic computational cellsize volume, and $\Delta t$ is the integration time-step. Similar to the approach in Ref.[62], it can be shown the magnetisation length distribution follows a Boltzmann probability distribution. For the 2-sublattice case, in general this distribution is a function of the magnetisation of both sub-lattices, $m_A$ and $m_B$, and is shown below for the isotropic case:

$$P_i(m_A, m_B) \propto m_i^2 \exp\left\{-\frac{M_S^0 V}{4\mu_i m_{e,i} k_B T}\left[\frac{(m_i^2 - m_{e,i}^2)^2}{m_{e,i}} \frac{(\mu_i + 3\tau_{ij} k_B T_N \tilde{\chi}_{\parallel,j})}{2\tilde{\chi}_{\parallel,i}} + \frac{(m_j^2 - m_{e,j}^2)}{m_{e,j}} 3\tau_{ij} k_B T_N m_i^2\right]\right\} \quad (15)$$

Verification of Equation (15) is given in the Supplementary Information.

The temperature is solved using a two-temperature model, where the electron and lattice temperature are coupled using rate equations as shown in Equation (16). The magnetisation is coupled to the electron bath via the Landau-Lifshitz damping, thus in the LLB equation we have $T = T_e$.

$$C_e \rho \frac{\partial T_e(\mathbf{r},t)}{\partial t} = \nabla . K \nabla T_e - G_e(T_e - T_l) + S$$
$$C_l \rho \frac{\partial T_l}{\partial t} = G_e(T_e - T_l) \quad (16)$$

In Equation (16) $C_e$ and $C_l$ are the electron and lattice specific heat capacities, $\rho$ is the mass density, $K$ is the thermal conductivity, and $G_e$ is the electron-lattice coupling constant, typically of the order $10^{18}$ W/m³K.

**Simulations.**

Simulations were performed using GPU-accelerated computations, using the finite difference formulation of the two-sublattice sLLB model coupled to the two-temperature model. The code used for this work is open-source and available at: https://github.com/SerbanL/Boris2. The cellsize was set to $1 \times 1 \times 2$ nm³ and thin films were simulated by employing periodic boundary conditions for the demagnetising field and differential operators. The magnetisation dynamics were solved using the Heun method with a time-step of 0.5 to 1 fs during the ultrafast demagnetisation stage and 1 fs to 5 fs during the longer magnetisation recovery stage. For the AFM thin film, material parameters were used as: $\alpha = 0.1$, $M_S^0 = 400$ kA/m, $A = 5$ pJ/m, $K_1 = 100$ kJ/m³ with easy axis perpendicular to the film, $A_h/a^3 = -10$ MJ/m³ with $a$ the lattice constant, $A_{nh} = -10$ pJ/m, $D = 1$ mJ/m², and $T_N = 500$ K. For the FM thin film we used: $\alpha = 0.1$, $M_S^0 = 600$ kA/m, $A = 10$ pJ/m, $K_1 = 380$ kJ/m³ with easy axis perpendicular to the film, $D = -1.5$ mJ/m², and $T_C = 500$ K. For the two-temperature model we used $C_e = 40$ J/kgK, $C_l = 130$ J/kgK, $K = 147$ W/mK, $\rho = 22650$ kg/m³.

# Emergence of Transient Domain Wall Skyrmions after Ultrafast Demagnetisation

Supplementary Information

Serban Lepadatu[1]

[1]*Jeremiah Horrocks Institute for Mathematics, Physics and Astronomy, University of Central Lancashire, Preston PR1 2HE, U.K.*

**Two-sublattice stochastic Landau-Lifshitz-Bloch model.**

Magnetisation dynamics in antiferromagnetic materials can be modelled using a 2-sublattice stochastic Landau-Lifshitz-Bloch (LLB) model. The corresponding LLB equation for classical ferromagnetic materials was given by Garanin[1], where spin-spin interactions are treated within the mean-field approximation (MFA), and was shown to agree well with predictions of atomistic simulations[2]. This was further used to model ultrafast laser-induced magnetisation dynamics together with a two-temperature model[3-5]. The stochastic form of the LLB (sLLB), which also takes into account thermal fluctuations, was given by Garanin and Chubykalo-Fesenko[6], with a second form later given by Evans et al.[7]. In particular, this second form was shown to reproduce the Boltzmann distribution predicted by the corresponding Fokker-Planck equation. A similar approach may be taken for ferrimagnetic and antiferromagnetic materials, as well as binary ferromagnetic alloys, by using a 2-sublattice MFA. A derivation of the ferrimagnetic 2-sublattice LLB equation was given by Atxitia et al.[8], which was later extended to cover the temperature range above the Curie temperature as well[9]. A stochastic form of the LLB equation was used recently to model sub-picosecond thermal pulses in antiferromagnetic FeRh[10], whilst a stochastic Landau-Lifshitz-Gilbert equation was used to model antiferromagnetic materials with the inclusion of interfacial DMI[11], and recently extended to include non-homogeneous inter-lattice exchange contributions[12]. The stochastic ferrimagnetic LLB equation was also used recently to study finite coarse-grained magnetic structures[13].

Here we show the 2-sublattice sLLB equation, based on the LLB equation from Refs.[8,9], applicable for antiferromagnetic, ferrimagnetic, as well as binary ferromagnetic alloys. We include both homogeneous and non-homogeneous inter-lattice exchange contributions, and recast the model in terms of accessible micromagnetic parameters, above and below the phase transition temperature. We also show the resulting magnetisation length probability distribution reproduces the expected 2-sublattice Boltzmann distribution throughout the temperature range, analogous to the sLLB equation for ferromagnets[7]. The explicit 2-sublattice sLLB equation is given in Equation (1) in terms of the macroscopic magnetisation, where we denote the 2 sublattices as $i = A, B$.

$$\frac{\partial \mathbf{M}_i}{\partial t} = -\tilde{\gamma}_i \mathbf{M}_i \times \mathbf{H}_{eff,i} - \tilde{\gamma}_i \frac{\tilde{\alpha}_{\perp,i}}{M_i} \mathbf{M}_i \times \left(\mathbf{M}_i \times \left(\mathbf{H}_{eff,i} + \mathbf{H}_{th,i}\right)\right)$$
$$+ \gamma_i \frac{\tilde{\alpha}_{\parallel,i}}{M_i} \left(\mathbf{M}_i \cdot \mathbf{H}_{\parallel,i}\right) \mathbf{M}_i + \mathbf{\eta}_{th,i} \quad (i = A, B)$$
(1)

The reduced gyromagnetic ratio is given by $\tilde{\gamma}_i = \gamma_i / (1 + \alpha_{\perp,i}^2)$, and the reduced transverse and longitudinal damping parameters by $\tilde{\alpha}_{\perp(\parallel),i} = \alpha_{\perp(\parallel),i} / m_i$, where $m_i(T) = M_i(T)/M_{S,i}^0$, with $M_{S,i}^0$ denoting the zero-temperature saturation magnetisation, and $Mi \equiv |\mathbf{M}i|$. The damping parameters are continuous at $T^N$ – the phase transition temperature – and given by:

$$\alpha_{\perp,i} = \alpha_i \left(1 - \frac{T}{3(\tau_i + \tau_{ij} m_{e,j}/m_{e,i})\tilde{T}_N}\right), \quad T < T_N$$
$$\alpha_{\parallel,i} = \alpha_i \left(\frac{2T}{3(\tau_i + \tau_{ij} m_{e,j}/m_{e,i})\tilde{T}_N}\right), \quad T < T_N$$
$$\alpha_{\perp,i} = \alpha_{\parallel,i} = \frac{2T}{3T_N}, \quad T \geq T_N$$
(2)

We denote $\tilde{T}_N$ the re-normalized transition temperature, given by:

$$\tilde{T}_N = \frac{2T_N}{\tau_A + \tau_B + \sqrt{(\tau_A - \tau_B)^2 + 4\tau_{AB}\tau_{BA}}}$$
(3)

The micromagnetic parameters $\tau_i$ and $\tau_{ij} \in [0, 1]$, are coupling parameters between exchange integrals and the phase transition temperature, such that $\tau_A + \tau_B = 1$ and $|J| = 3\tau k_B T_N$. Here $J$ is the exchange integral for intra-lattice ($i = A,B$) and inter-lattice ($i,j = A,B, i \neq j$) coupling respectively. For a simple antiferromagnet we have $\tau_A = \tau_B = \tau_{AB} = \tau_{BA} = 0.5$. The normalised equilibrium magnetisation functions $m_{e,i}$ are obtained from the Curie-Weiss law as:

$$m_{e,i} = B\left[\left(m_{e,i}\tau_i + m_{e,j}\tau_{ij}\right)3\tilde{T}_N/T + \mu_i\mu_0 H_{ext}/k_B T\right], \tag{4}$$

where $B(x) = \coth(x) - 1/x$, and $\mu_i$ is the atomic magnetic moment. The magnetisation length is not constant, and can differ from the equilibrium magnetisation length, giving rise to a longitudinal relaxation field which includes both intra-lattice and inter-lattice contributions:

$$\mathbf{H}_{\|,i} = \left\{\frac{1}{2\mu_0 \tilde{\chi}_{\|,i}}\left(1 - \frac{m_i^2}{m_{e,i}^2}\right) + \frac{3\tau_{ij}k_B T_N}{2\mu_0 \mu_i}\left[\frac{\tilde{\chi}_{\|,j}}{\tilde{\chi}_{\|,i}}\left(1 - \frac{m_i^2}{m_{e,i}^2}\right) - \frac{m_{e,j}}{m_{e,i}}\left(\hat{\mathbf{m}}_i \cdot \hat{\mathbf{m}}_j\right)\left(1 - \frac{m_j^2}{m_{e,j}^2}\right)\right]\right\}\mathbf{m}_i, \quad T < T_N$$

$$\mathbf{H}_{\|,i} = -\left\{\frac{1}{\mu_0 \tilde{\chi}_{\|,i}} + \frac{3\tau_{ij}k_B T_N}{\mu_0 \mu_i}\left[\frac{\tilde{\chi}_{\|,j}}{\tilde{\chi}_{\|,i}} - \frac{m_{e,j}}{m_{e,i}}\left(\hat{\mathbf{m}}_i \cdot \hat{\mathbf{m}}_j\right)\right]\right\}\mathbf{m}_i, \quad T > T_N \tag{5}$$

Here $\hat{\mathbf{m}}_i = \mathbf{m}_i / m_i$, and the relative longitudinal susceptibility is $\tilde{\chi}_{\|,i} = \chi_{\|,i} / \mu_0 M_{S,i}^0$, where:

$$k_B T \chi_{\|,i} = \frac{\mu_i B_i'\left(1 - 3\tau_j \tilde{T}_N B_j'/T\right) + \mu_j 3\tau_{ij}\tilde{T}_N B_i' B_j'/T}{\left(1 - 3\tau_i \tilde{T}_N B_i'/T\right)\left(1 - 3\tau_j \tilde{T}_N B_j'/T\right) - \tau_{ij}\tau_{ji} B_i' B_j'\left(3\tilde{T}_N/T\right)^2}, \tag{6}$$

and $B_i' \equiv B'_{m_{e,i}}\left[\left(m_{e,i}\tau_i + m_{e,j}\tau_{ij}\right)3\tilde{T}_N/T\right]$.

The effective field in Equation (1) is a sum of all the interaction fields, and given by $\mathbf{H}_{eff,i} = \mathbf{H}_{ext,i} + \mathbf{H}_{demag,i} + \mathbf{H}_{ani,i} + \mathbf{H}_{ex,i}$. In particular $\mathbf{H}_{demag,i}$ is the demagnetising field calculated for the net magnetization $(\mathbf{M}_A + \mathbf{M}_B)/2$, and applied equally to both sub-lattices. The uniaxial anisotropy field is given by:

$$\mathbf{H}_{ani,i} = \frac{2K_{1,i}}{\mu_0 M_{e,i}^2}\left(\mathbf{M}_i \cdot \mathbf{e}_A\right)\mathbf{e}_A \tag{7}$$

Here $\mathbf{e}_A$ is the symmetry axis, $M_{e,i} = m_{e,i}M_{S,i}^0$, and $K_{1,i}$ follows the temperature dependence $K_{1,i} = K_{1,i}^0 m_{e,i}^3$. The exchange field includes both the isotropic direct exchange term, as well as the interfacial DMI exchange term. The direct exchange term includes the usual intra-lattice contribution, as well as homogeneous and non-homogeneous inter-lattice contributions, and is given by:

$$\mathbf{H}_{ex,i} = \frac{2A_i}{\mu_0 M_{e,i}^2}\nabla^2\mathbf{M}_i + \frac{4A_{h,i}}{\mu_0 M_{e,i}M_{e,j}}\mathbf{M}_j + \frac{A_{nh,i}}{\mu_0 M_{e,i}M_{e,j}}\nabla^2\mathbf{M}_j \tag{8}$$

The intra-lattice exchange stiffness $A_i$ has the temperature dependence $A_i = A_i^0 m_{e,i}^2$, whilst the inter-lattice exchange stiffnesses have the temperature dependences $A_{h(nh),i} = A_{h(nh),i}^0 m_{e,i} m_{e,j}$. The interfacial DMI exchange field is given by, applicable for systems with $C_{nv}$ symmetry:

$$\mathbf{H}_{iDMI,i} = -\frac{2D_i}{\mu_0 M_{e,i}^2}\left(\frac{\partial M_z}{\partial x}, \frac{\partial M_z}{\partial y}, -\frac{\partial M_x}{\partial x} - \frac{\partial M_y}{\partial y}\right), \tag{9}$$

with the DMI exchange parameter having the temperature dependence $D_i = D_i^0 m_{e,i}^2$.

Finally, the terms $\mathbf{H}_{th,i}$ and $\mathbf{\eta}_{th,i}$ are stochastic quantities with zero spatial, vector components, and inter-lattice correlations, and whose components follow Gaussian distributions with zero mean and standard deviations given respectively by:

$$\begin{aligned}H_{th,i}^{std.} &= \frac{1}{\alpha_{\perp,i}}\sqrt{\frac{2k_B T(\alpha_{\perp,i} - \alpha_{\parallel,i})}{\gamma_i \mu_0 M_{S,i}^0 V \Delta t}} \\ \eta_{th,i}^{std.} &= \sqrt{\frac{2k_B T \alpha_{\parallel,i}\gamma_i M_{S,i}^0}{\mu_0 V \Delta t}}\end{aligned} \tag{10}$$

Here $V$ is the stochastic computational cellsize volume, and $\Delta t$ is the integration time-step. Similarly to the approach in Ref.[7], it can be shown the magnetisation length distribution follows a Boltzmann probability distribution. For the 2-sublattice case, in general this distribution is a function of the magnetisation of both sub-lattices, $m_A$ and $m_B$, and is shown below for the isotropic case.

$$P_i(m_A, m_B) \propto m_i^2 \exp\left\{-\frac{M_S^0 V}{4\mu_i m_{e,i} k_B T}\left[\frac{(m_i^2 - m_{e,i}^2)^2}{m_{e,i}}\frac{(\mu_i + 3\tau_{ij}k_B T_N \tilde{\chi}_{\parallel,j})}{2\tilde{\chi}_{\parallel,i}} + \frac{(m_j^2 - m_{e,j}^2)}{m_{e,j}}3\tau_{ij}k_B T_N m_i^2\right]\right\} \tag{11}$$

Thus a good test of the 2-sublattice sLLB equation is to compute the magnetisation length histograms and compare the numerical results with the analytical predictions of Equation (11). To this end, a simplification may be used by taking a profile through the bi-variate probability distribution for $m_j \cong m_{e,j}$, which eliminates the dependence on $m_j$. This is shown in Figure S1(a) for the antiferromagnetic case, where very good agreement is observed between numerical and analytical results, even when very close to the Néel temperature. The equilibrium magnetisation length and longitudinal susceptibility obtained from the MFA is shown in Figure S1(b), where we also test the mean magnetisation length obtained from numerical simulations reproduces the input magnetisation function even in the vicinity of the Néel temperature. The LLB equation in the MFA is applicable effectively for an infinite system, however when used for magnetic thin films, particularly with periodic boundary conditions as is the case in this work, the approximation holds well. Finally it should be noted the sLLB equation shown here reduces to the ferromagnetic sLLB equation by setting $\tau_A = 1$ and $\tau_B = \tau_{AB} = \tau_{BA} = 0$.

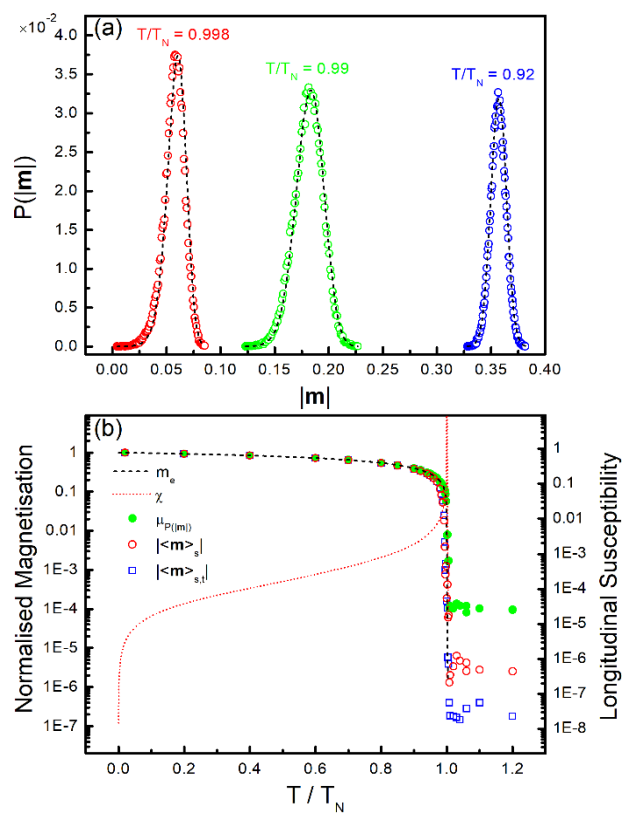

**Figure S1. Antiferromagnetic sLLB Equation Properties.** (a) Computed probability distribution of magnetisation length for different temperatures on sub-lattice A, sampled for $m_B \cong m_{e,B}$. The dashed lines show the predicted Boltzmann distributions using Equation (11). (b) Computed normalised magnetisation length, compared with the equilibrium magnetisation function obtained from the MFA (dashed line). Solid circles show the magnetisation length probability distribution mean, empty circles are obtained by spatial averaging, and empty squares are obtained by both spatial and time averaging. The dotted line shows the longitudinal susceptibility.

Studies of ultrafast magnetisation dynamics have revealed a large difference between electron and spin dynamics on time-scales of the order 1 picosecond and below, explained in terms of a 3-temperature model which includes the electron, spin, and lattice temperatures[14], and later formulated as a microscopic 3-temperature model[15]. This latter approach was shown to be equivalent to an LLB formulation[4] which accounts for the different electron and lattice temperatures on ultra-short time-scales. Within this formulation the phonon energy is absorbed by the delocalized electrons, which are coupled to the lattice electrons via rate equations as shown in Equation (12). The magnetisation is coupled to the electron bath via the Landau-Lifshitz damping, thus in the MFA formulation of the LLB equation we have $T = T_e$. The approach based on the LLB equation is able to reproduce both the ultrafast demagnetisation due to a heat pulse, and the subsequent magnetisation recovery, as well as the longer time-scale magnetisation processes.

$$C_e \rho \frac{\partial T_e(\mathbf{r},t)}{\partial t} = \nabla . K \nabla T_e - G_e(T_e - T_l) + S$$
$$C_l \rho \frac{\partial T_l}{\partial t} = G_e(T_e - T_l)$$
(12)

In Equation (12) $C_e$ and $C_l$ are the electron and lattice specific heat capacities, $\rho$ is the mass density, $K$ is the thermal conductivity, and $G_e$ is the electron-lattice coupling constant, typically of the order $10^{18}$ W/m$^3$K. Here the heat source S due to the laser pulse is assumed to have Gaussian spatial and temporal profiles given by:

$$S = P_0 \exp\left(\frac{-|\mathbf{r}-\mathbf{r}_0|}{d^2/4\ln(2)}\right) \exp\left(\frac{-(t-t_0)^2}{t_R^2/4\ln(2)}\right) \quad (W/m^3)$$
(13)

where $d$ and $t_R$ are full-width at half-maximum (FWHM) values describing the laser pulse spatial and time-dependence.

**Antiferromagnetic skyrmion creation with varying DMI strength.**

The same calculations given in the main paper are repeated here for varying DMI strength. Results are shown in Figure S2, where the same *r* fitting factors are used ($r_+ \cong 0.3$ and $r_- \cong 0.4$) for the model in Equation (14). The region for skyrmion stability in this case is in the range D = 0.8 – 1.0 mJ/m², with the smaller D values resulting in skyrmions with decreased thermal stability. Thus to maintain an average skyrmion creation rate of 1 we require larger $A_{S\pm}^{AFM}$ values, ranging from $A_{S-}^{AFM} = 0.01$ µm² and $A_{S+}^{AFM} = 0.15$ µm² for D = 1.0 mJ/m², to $A_{S-}^{AFM} = 0.02$ µm² and $A_{S+}^{AFM} = 0.4$ µm² for D = 0.8 mJ/m². A reasonable agreement is obtained between the numerical results with statistical information extracted from up to 50 events for each set of simulation parameters, and the simple phenomenological model in Equation (14).

$$N_{S\pm}^{AFM} = \frac{A_{PM}}{A_{S\pm}^{AFM}}\left[1 + r_\pm\left(N_{S\pm}^{AFM} - 1\right)\right]^{\pm 1} \tag{14}$$

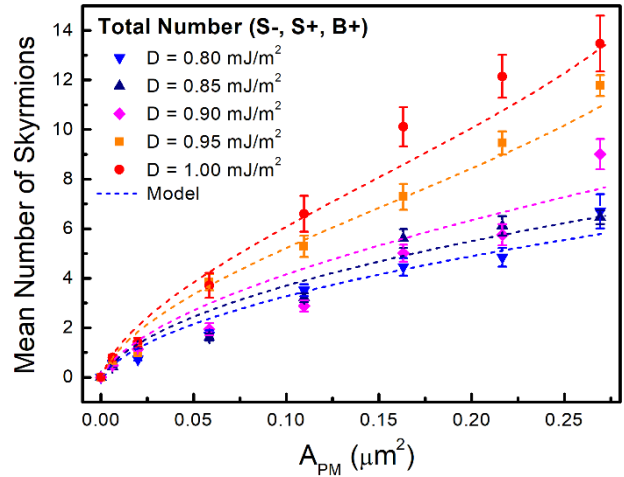

**Figure S2. Effect of DMI on Skyrmion Creation Rates.** Total number of topological objects, S-, S+, and B+ for DMI strength ranging from 0.8 mJ/m2 up to 1.0 mJ/m2. The dashed lines are obtained using the model in Equation (14).